\documentclass[a4paper]{article}

\usepackage{19lomcon}        
\usepackage{cite}             
\usepackage{epsfig}           
\usepackage{epstopdf}

\begin{document}


\title{THE RADIO SYNCHROTRON BACKGROUND \\ A COSMIC CONUNDRUM\\ {\rm \small (to appear in ``{\it Particle Physics at the Year of the 150th Anniversary of Mendeleev's Periodic Table of Chemical Elements},'' A. Studenikin, Ed.)} \\ {\rm \small Proceedings of the 19th Lomonosov Conference on Elementary Particle Physics} }

\author{Jack Singal \email{jsingal@richmond.edu}
}

\affiliation{University of Richmond, Richmond, Virginia, USA}


\date{}
\maketitle


\begin{abstract}

It has recently become apparent that the background level of diffuse radio emission on the sky is significantly higher than the level that can result from known extragalactic radio source classes or our Galaxy given our current understanding of its large-scale structure.~ In contrast to the more well-known and well-constrained cosmological and astrophysical backgrounds at microwave, infrared, optical/UV, X-ray, and gamma-ray wavelengths, this ``radio synchrotron background'' at radio wavelengths provides clear motivation for considering the possibilities of new astrophysical sources and new particle-based emission mechanisms.

\end{abstract}
\vspace{-0.1in}
\section{Radio Background Observations}\label{s1}
\begin{figure}[t]
  \begin{minipage}{5cm}
\centering
\includegraphics[height=4cm]{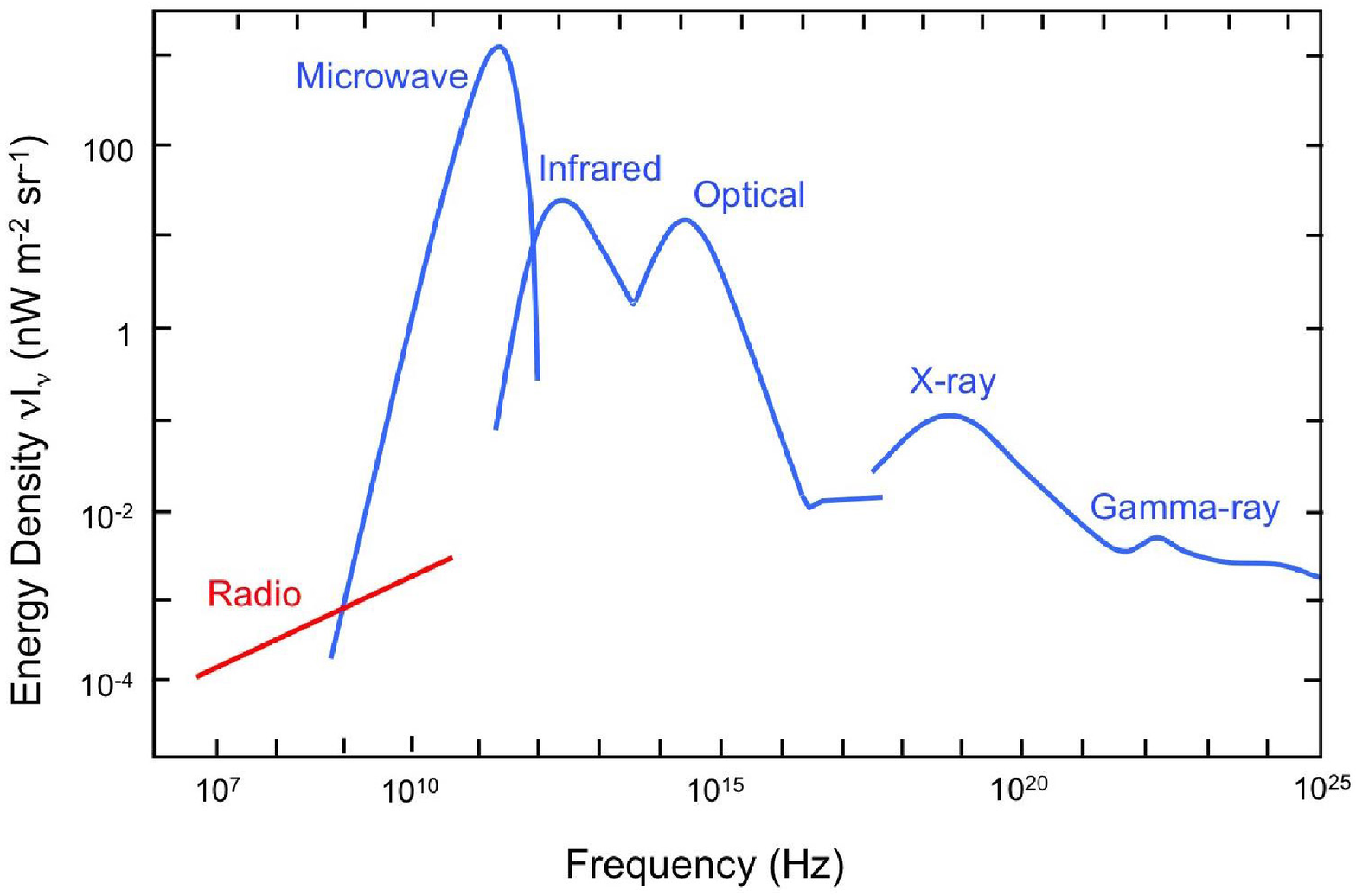}
  \end{minipage}
%
\hfill
  \begin{minipage}{5cm}
     \centering
     \includegraphics[height=5cm]{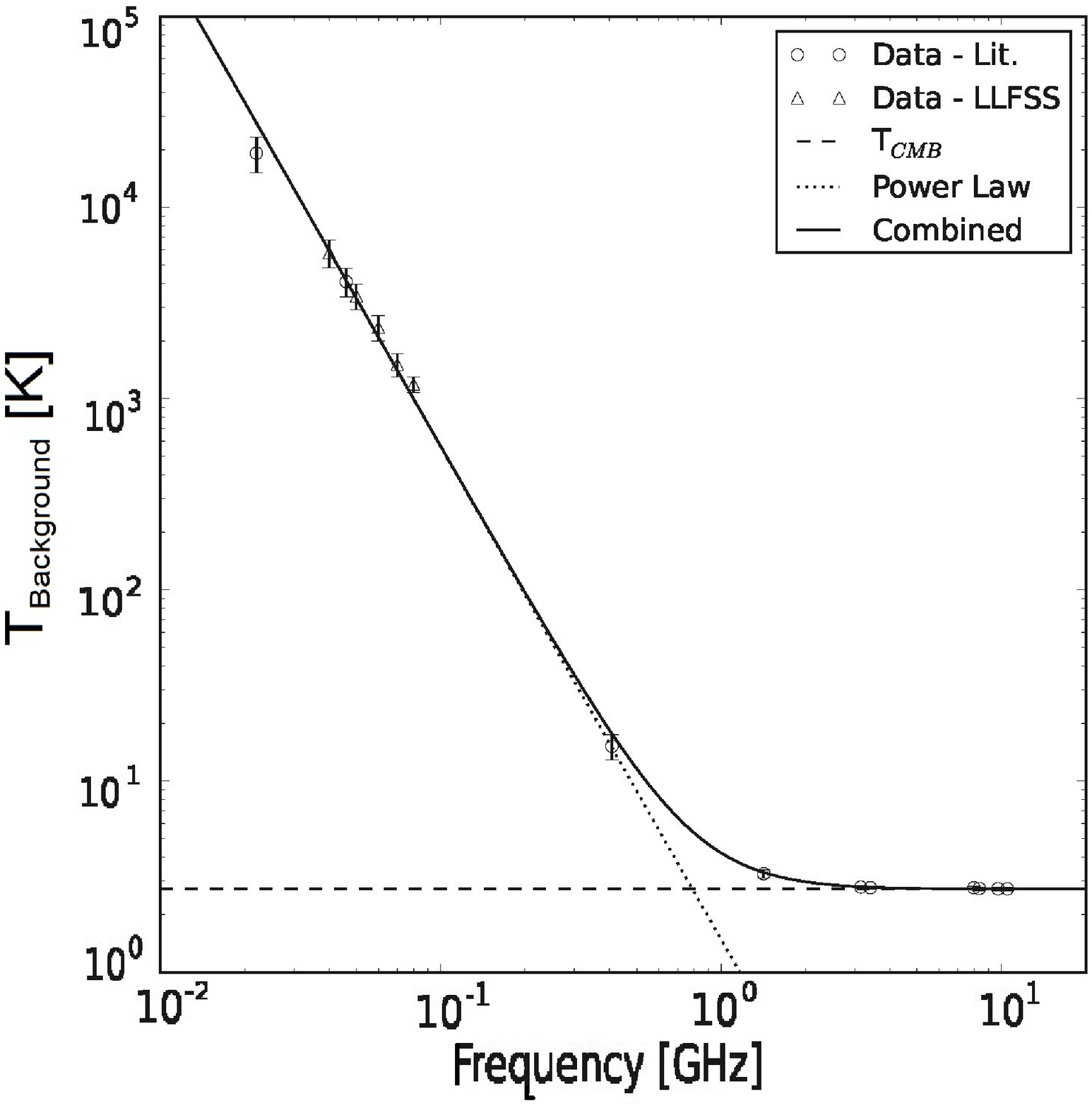}
  \end{minipage}
\hfill
     \centering
     \includegraphics[width=7cm]{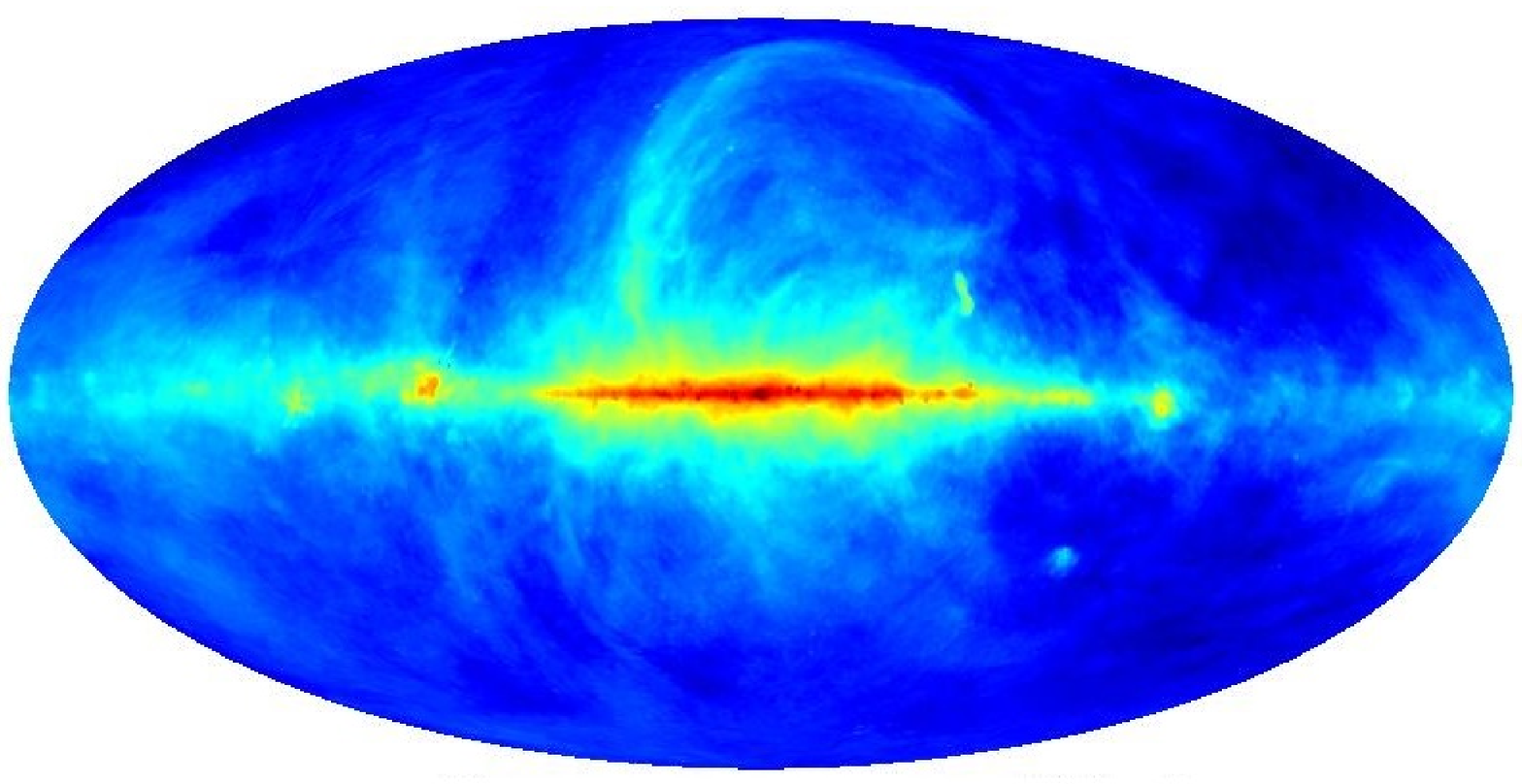}
\caption{ {\bf Left}: Spectral energy surface brightness density of the photon backgrounds in the Universe.~ The radio synchrontron background dominates at MHz frequencies as discussed in \S \ref{s1}.~ {\bf Right}: The measured background brightness spectrum in radiometric temperature units reproduced from \cite{DT18} and \cite{CP}, as measured by several different instruments or surveys including that work.~ {\bf Bottom}: The famous Haslam map \cite{Haslam} from the 1980s shows the spatial structure of diffuse astrophysical radio emission on the sky.~ It is dominated by the Galactic plane-parallel structure.~ Only near the Galactic poles is the emission dominated by the isotropic background component which is the result of some as of now unknown combination of numerous extragalactic sources and/or a large halo surrounding our Galaxy.~ }
\label{f1}
\end{figure}

\vspace{-0.1in}
In astrophysics, the concept of a `photon background' refers to the level of diffuse light in a certain waveband that is not attributed to resolved individual sources and from which the large-scale spatially-varying components --- such as associated with the Galactic plane structure or the doppler dipole --- have been subtracted or masked.~ The photon backgrounds include most famously the microwave background, which originates from the last scattering surface in the early Universe, but also backgrounds in the infrared, optical/UV, X-ray, and gamma-ray bands which are due to the combined surface brightness of many extragalactic sources such as star-forming and active galaxies.\footnote{Because the amount of light that remains diffuse --- i.e.~not resolved into individual sources --- depends on the resolution of any particular observation, the level ascribed to each of these photon backgrounds can vary definitionally.~ To achieve consistency in this regard, we adopt the premise used by the {\it Fermi} Gamma-ray space telescope collaboration (e.g.~\cite{Fermi1}) which considers a photon background to be all of the integrated surface brightness in the particular waveband after removal of large-scale spatially varying structure.}  The surface brightness level of all of the photon backgrounds is shown in the left panel of Figure \ref{f1}.~

Determination of the level of the photon background at radio wavelengths is complicated by a number of factors.~ Because of the general inverse scaling of resolution with wavelength, radio observations are often done with multi-telescope interferometer arrays which are insensitive to the absolute brightness level being observed and can only measure {\it differences} in brightness.~ Radio observations done with one telescope are in principle sensitive to the overall absolute brightness level but this is often complicated by spillover pickup from other points on the sky or ground, atmospheric emission, radio frequency interference, and uncertainty in instrumental system noise contributions.~ Most radio observations thus lack what is termed an absolute zero-level calibration rendering them insensitive to the photon background at radio wavelengths.

Some of the few measurements that have been able to meaningfully discern the level of the radio background are shown in the right panel of Figure \ref{f1}.  They are those from the ARCADE~2 experiment \cite{Fixsen11,Singal11} above $\nu =3 \mathrm{~GHz}$, \cite{RR86} at 1.4~GHz, \cite{Haslam} at 408~MHz, \cite{Maeda99} at 45~MHz, \cite{Roger99} at 22~MHz, and \cite{DT18} at the remaining frequencies plotted in that figure.  Of these, only the ARCADE~2 measurements had an absolute zero-level calibration as a primary goal.~ However combining the ARCADE~2 results with the others capable of measuring the background revealed a suprisingly high level, with the radiometric temperature ($T_\mathrm{B}$) as a function of frequency ($\nu$) fitting the form (e.g.~\cite{CP}):
\begin{equation}\label{eq1}
T_\mathrm{B} ({\mathrm K}) = 24.1 \pm 2.1\,  
\Biggl(\frac{\nu}{310 \mathrm{~MHz}}\Biggr)^{-2.6 \pm 0.04}
\end{equation}
The power-law scaling of -2.6 is indicative of synchrotron radiation, which is indeed the dominant mechanism for radio emission in the Universe.~ Synchrotron radiation results from high energy relativistic charged particles (usually electrons) spiraling in magnetic fields.~ The level of synchrotron emission from a given region is thus dependent on both the magnetic field energy density and the electron cosmic ray energy density in that region and so is a probe of both (e.g.~\cite{RL79}).~ The synchrotron spectrum has led to the surprisingly high level of radio background emission being termed the ``radio synchrotron background'' which will hereafter be abbreviated RSB.
\vspace{-0.1in}
\section{Radio Background Source Mysteries}\label{s2}
\begin{figure}[t]
  \begin{minipage}{5cm}
\centering
\includegraphics[width=7.5cm]{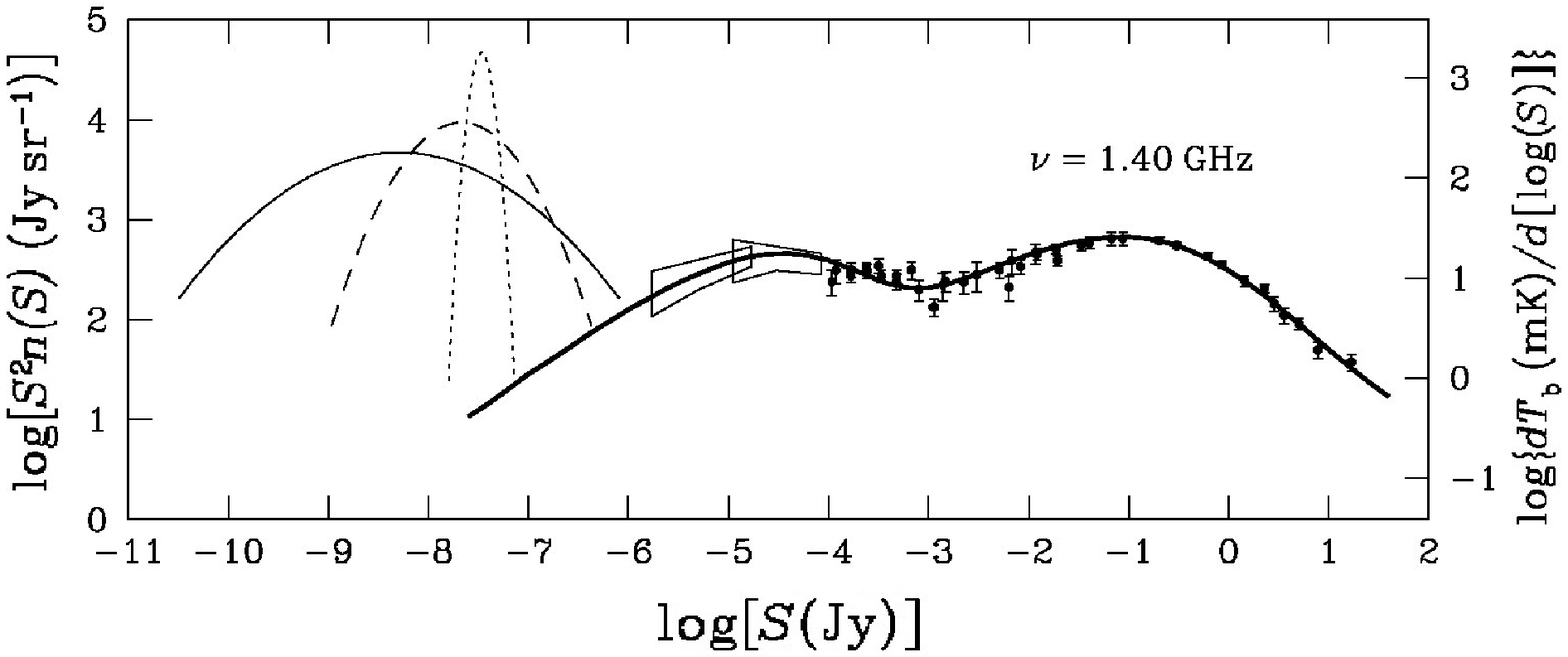}
  \end{minipage}
%
\hfill
  \begin{minipage}{5cm}
     \centering
     \includegraphics[height=4cm]{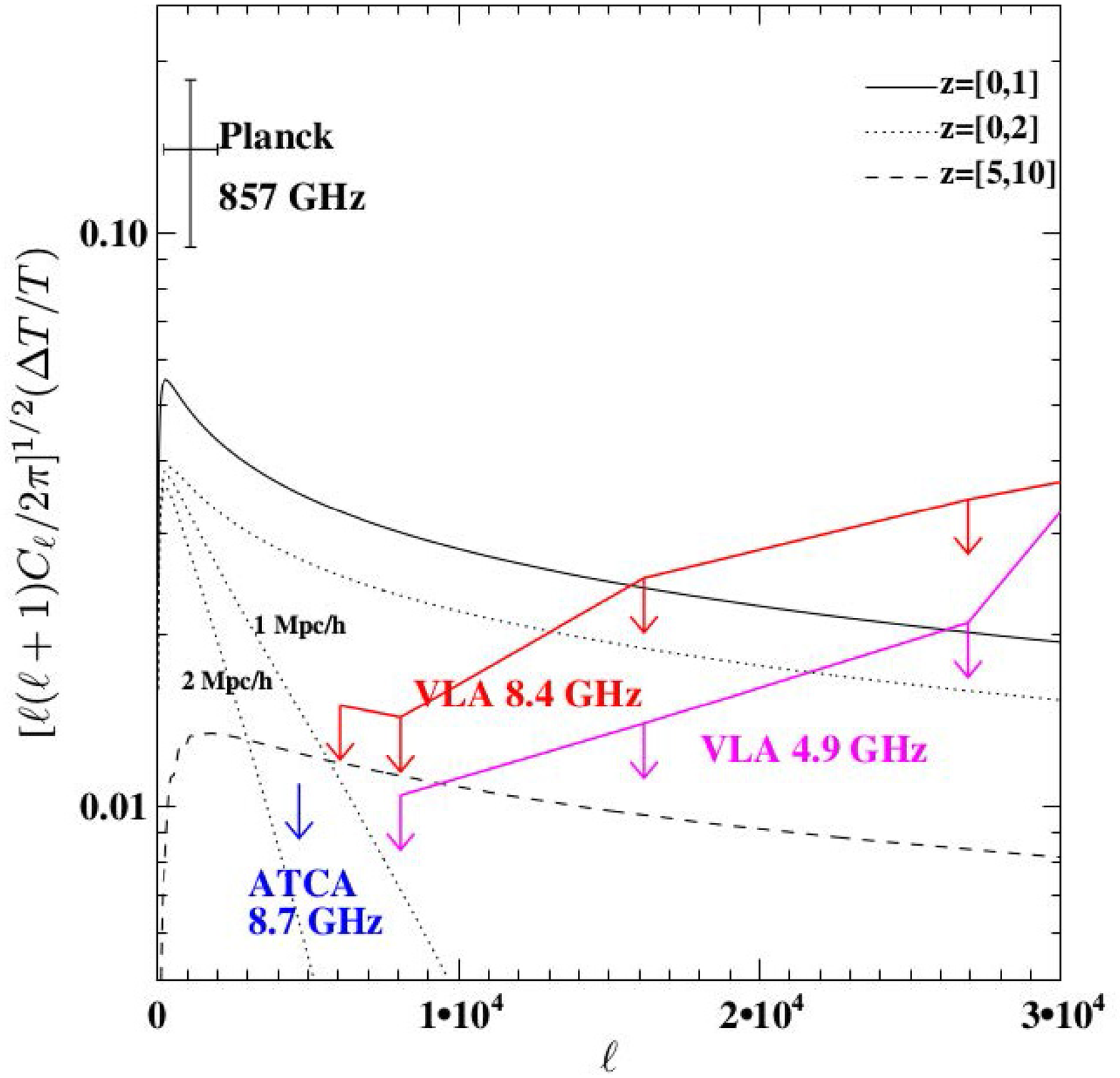}
  \end{minipage}

\caption{ {\bf Left}: Reproduced from \cite{Condon12} and \cite{CP}.~The measured radio source counts at 1.4~GHz (differential number of sources versus flux) shown by the heavy curve limit their contribution to only around one-fifth of the reported RSB brightness.~ The three parabolas at nano-Jansky levels represent hypothetical new source populations that can produce the reported background surface brightness, indicating that if a heretofore unknown low-flux source class is responsible it must be extremely numerous as discussed in \S \ref{s2}.~ {\bf Right}: Reproduced from \cite{Holder14} and \cite{CP}.~ Expected angular power from clustering for several ranges in redshift for the contributions to the unresolved radio background, as well as the currently observed upper limits on angular power from incidental measurements.~ For the redshift interval $z$=0--2 (dotted), the effect of each source being extended is shown: top to bottom are FWHM$_{\rm smooth}$ = 0, 1, 2$h^{-1}$ comoving Mpc.~ These upper bounds indicate that the source population for the RSB must not be point sources clustered like the large scale structure of the Universe.~ }
\label{f2}
\end{figure}

\vspace{-0.1in}
The reported bright RSB level is now in extreme tension with the lower total surface brightness that the integrated contribution of known astrophysical source populations and emission mechanisms can produce, as reviewed recently in \cite{CP}.~ The deepest radio source counts (tallies of the number of radio sources versus flux) limit the integrated surface brightness from known classes of extragalactic radio sources to only $\sim 1/5$ of the reported radio background brightness level.~ A summary of our knowledge of radio source counts is shown in the left panel of Figure~\ref{f2}.~ One could ask if a class or classes of heretofore unknown radio sources could produce the observed RSB level.~ In order to be currently unobserved, these sources would have to be low flux and therefore incredibly numerous \cite{Condon12}, exceeding by orders of magnitude the total number of galaxies in the observable Universe.

Furthermore, the observed far-infrared (FIR) background would be exceeded if the bright radio background were produced by sources that follow the known FIR/radio correlation of star-forming galaxies (e.g.~\cite{YL12}), and arcminute-scale fluctuations in the background are so small that any potential sources tracing the large scale structure of the universe must be either very diffuse or at high redshifts \cite{Holder14} as seen in the right panel of Figure \ref{f2}.~ Diffuse extragalactic sources such as cluster mergers (e.g.~\cite{FL15}) have been proposed as an alternative, but these present some challenges as well \cite{Vernstrom17}.

A large, bright, roughly spherical synchrotron halo surrounding our Galaxy could explain part of the RSB level \cite{SC13}, and a large halo is consistent with some cosmic-ray propagation models (e.g.~\cite{OS13}).~ However: (i) In the weak ($B \sim$1\,$\mu$G) halo magnetic field inferred from Faraday rotation measures of extragalactic sources (e.g.~\cite{Taylor02}), the same cosmic-ray electrons needed to produce the synchrotron halo would overproduce the observed X-ray background via inverse-Compton emission \cite{RB1}, (ii) such a large, bright halo would make our Galaxy wholly unique among nearby spiral galaxies \cite{RB1} and (iii) the Galactic latitude- and longitude-dependent structure of the diffuse radio emission as inferred from the presently available maps at several frequencies does not support the presence of such a large, round halo \cite{Fornengo14}.~ 

Also, one must consider the possible implications for the observed level of the global (sky-averaged) 21-cm signature from reionization.~ The 21-cm absorption feature results from the lower temperature of the IGM relative to the CMB (at high redshifts).~ Any RSB at those redshifts increases the effective temperature of the CMB --- increasing the depth of the 21-cm absorption trough.~ Several works \cite{EW18,FH18,MF19} have calculated the effects of any RSB present at high redshifts on the trough depth to be quite significant, possibly exceeding current observational constraints on the trough depth.~ This indicates that most of the RSB should originate from after the reionization era.

\vspace{-0.1in}
\section{Hypothesized Source Classes}\label{s3}
\vspace{-0.1in}
The mystery of the RSB presents unique opportunities for considering new particle physics, in addition to reconsidering our understanding of extragalactic radio source classes and the halo emission structure of our Galaxy.~ A number of potential particle-related explanations for RSB emission mechanisms have been suggested, including intergalactic dark matter decays and annihilations in halos and filaments (e.g.~\cite{Fornengo11,FL14,Hooper14}), ultracompact halos \cite{Yang13}, or in the relatively early Universe (e.g.~\cite{CV14}), ``dark stars'' in the early universe \cite{Spolyar09}, supernovae of massive population~III (first generation) stars \cite{Biermann14}, and dense nuggets of quarks \cite{LZ13}.
\vspace{-0.1in}
\section{Discussion and Future Prospects}
\vspace{-0.1in}
There is (almost certainly) a significantly brighter radio synchrotron background than can be accounted for by known extragalactic source classes and most models of Galactic diffuse emission.~ Whatever the source population(s) of the RSB is/are, they are highly constrained.~ As discussed in \S \ref{s2}, if they are extragalactic, the new source population(s) would have to: I) be incredibly numerous, II) not follow the radio/far-infrared correlation, III) have high magnetic fields so as not to overproduce observed X-ray brightnesses via inverse-Compton, and IV) be at least one of: a) high (but not too high) redshift, b) not clustered like typical large scale structure, and/or c) diffuse.~If the RSB originates from our Galaxy, it would make our Galaxy highly anomalous.~ 

Observational projects are planned and in progress which will further our understanding of the RSB.~ One such effort is an absolutely calibrated zero-level map at 310~MHz utilizing the Green Bank Telescope and custom instrumentation.~ This will be the first ever large-scale map of the diffuse radio emission which has an absolute zero-level calibration as a primary goal.~ The project is currently approved for 24~hours of observing which will result in a porous map, while a proposal is being prepared to extend this for a long-term observing campaign to make a full-sky map, essentially updating the 40~year old diffuse radio map shown in the bottom panel of Figure \ref{f1}.~ Another approved project is 8~hours of observing with the LOFAR interferometer array to search for arcminute-scale anisotropies in the RSB (or limits thereof) at the MHz frequencies where it dominates rather than relying on legacy observations at GHz frequencies as in the right panel of Figure \ref{f2}.~ These observations will refine the task of proposing hypotheses to explain the origin of the RSB.

\section*{Acknowledgments}

J.~Singal thanks the many people involved in the RSB community, including A.~Kogut, D.~Fixsen, M.~Seiffert, M.~Limon, E.~Wollack, J.~Condon, R.~Bradley, S.~Srikanth, K.~Kellermann, E.~Murphy, B.~Nahn, D.~Bordeneave, K.~Makhija, S.~Horiuchi, D.~Lucero, A.~Ofringa, N.~Fornengo, M.~Regis, G.~Holder, D.~Ballantyne, J.~Dowell, B.~Harms, P.~Biermann, T.~Linden, R.~Monsalve, P.~Mertsch, D.~Scott, T.~Vernstrom, L.~Xu, T.~Bunn, E.~Jones, J.~Haider, V.~Petrosian \& R.~Subrahmanyan. J.~Singal also thanks the organizers of the 19th Lomonosov Conference.


\end{document}